\documentclass[runningheads]{llncs}
\usepackage[T1]{fontenc}
\usepackage{graphicx}
\usepackage[english]{babel}
\usepackage{booktabs}
\usepackage{textcomp}
\usepackage{xcolor}
\usepackage[font={small}]{caption}
\usepackage{float}
\usepackage{adjustbox}
\usepackage{enumitem}
\usepackage{todonotes}
\usepackage[many]{tcolorbox}
\usepackage{cite}
\usepackage{hyperref}
\newtcolorbox{colorb}{
enhanced,
boxrule=0pt,frame hidden,
borderline west={2pt}{0pt}{green!50!black},
colback=green!05!white,
sharp corners
}
\newtcolorbox{colora}{
enhanced,
boxrule=0pt,frame hidden,
borderline west={2pt}{0pt}{gray!50!black},
colback=gray!05!white,
sharp corners
}

\begin{document}
\title{Regulatory Requirements Engineering in Large Enterprises: An Interview Study on the European Accessibility Act}
\titlerunning{Regulatory Requirements Engineering in Large Enterprises}
%
\author{Oleksandr Kosenkov\inst{1,2}\and
Michael Unterkalmsteiner\inst{1} \and
Daniel Mendez\inst{1,2} \and
Jannik Fischbach\inst{2,3}}

\authorrunning{O. Kosenkov et al.}

\institute{Blekinge Institute of Technology, Karlskrona, Sweden\\ \email{firstname.lastname@bth.se} \and
fortiss GmbH, Munich, Germany\\
\email{lastname@fortiss.org} \and
Netlight Consulting GmbH, Munich, Germany\\
\email{jannik.fischbach@netlight.com}}

\maketitle              
\begin{abstract}
Context: Regulations, such as the European Accessibility Act (EAA), impact the engineering of software products and services. Managing that impact while providing meaningful inputs to development teams is one of the emerging requirements engineering (RE) challenges.

Problem: Enterprises conduct Regulatory Impact Analysis (RIA) to consider the effects of regulations on software products offered and formulate requirements at an enterprise level. Despite its practical relevance, we are unaware of any studies on this large-scale regulatory RE process.

Methodology: We conducted an exploratory interview study of RIA in three large enterprises. We focused on how they conduct RIA, emphasizing cross-functional interactions, and using the EAA as an example.

Results: RIA, as a regulatory RE process, is conducted to address the needs of executive management and central functions. It involves coordination between different functions and levels of enterprise hierarchy. Enterprises use artifacts to support interpretation and communication of the results of RIA. Challenges to RIA are mainly related to the execution of such coordination and managing the knowledge involved.

Conclusion: RIA in large enterprises demands close coordination of multiple stakeholders and roles. Applying interpretation and compliance artifacts is one approach to support such coordination. However, there are no established practices for creating and managing such artifacts.

\end{abstract}
\section{Introduction}\label{sec:introduction}
\textbf{Context} 
In recent years, research on regulations as a source of requirements for software-intensive products and services (SIPS) has received much attention~\cite{mubarkoot2023software, usman2020compliance}. Existing requirements engineering (RE) studies identified multiple challenges to processing regulations and implementing regulatory compliance. Large enterprises are often considered well-positioned to address the challenges related to the demand for legal expertise and additional resources for processing regulations~\cite{hjerppe2019general}, however large enterprises may face other kinds of challenges. 

\textbf{Problem} One recent regulation with a broad impact on the development of software-intensive products and services is the European Accessibility Act (EAA). EAA introduces requirements on products and services to make them accessible for persons with disabilities, and its implementation is related to multiple challenges~\cite{bittenbinder2023responsibilities}. An increasing number of enterprises conduct Regulatory Impact Analysis (RIA) as the first step in their enterprise-wide regulatory RE. Still, there are no empirical studies that would explore how RIA is executed in practice and the way compliance stakeholders interact. There is also a lack of studies that consider executive management as an important group of stakeholders in regulatory RE.

\textbf{Research Goal} We conduct an exploratory interview study in three large enterprises to answer the following research questions (RQ):

\begin{enumerate}[label=\bfseries RQ \arabic*:,leftmargin=*,labelindent=0em]
    \item How do large enterprises conduct Regulatory Impact Analysis (RIA) for the EAA?
    \item What are the challenges related to RIA?
    \item How do large enterprises approach cross-functional engineering-legal interaction in the process of RIA?
\end{enumerate}

\textbf{Contribution} We shed light on (1) the goals and structure of the RIA process, (2) challenges to the RIA process in practice, and (3) the role and approaches to cross-functional coordination and knowledge management in RIA.

\textbf{Outline} The rest of the paper is structured as follows. In Section~\ref{sec:background}, we introduce concepts and terms used in this study. Section~\ref{sec:relatedWork} provides an overview of the related studies. We outline the methodology used to execute this study in Section~\ref{sec:methodology}. In Section~\ref{sec:results}, we introduce our research results by summarizing the aspects and challenges of the RIA process. We provide a synthesis and discussion of our results in Section~\ref{sec:discussion} and present our conclusion and future research plans in Section~\ref{sec:conclusion}.

\section{Background}\label{sec:background}
Enterprises developing \textit{software-intensive products and services (SIPS)} for internal use or as products need to implement a verifiable state of conformance of such systems to applicable legal norms (\textit{legal compliance}). Regulations (i.e., public, general, obligatory sources of norms issued by regulators) are sources of important legal requirements due to the financial or other penalties enterprises can face for non-compliance. Large enterprises usually have different levels of governance, constituting the context for the SIPS they develop. \textit{Executive management} is the highest management level in enterprises and can include roles such as Chief Executive Officer or Chief Information Officer. Enterprises also include different \textit{central functions} (e.g., legal, risk management). The structure of large enterprises includes \textit{organizational units (OUs)} responsible for different geographical areas and/or types of products in the company. In enterprises, OUs include multiple teams developing software-intensive services and products - \textit{SIPS teams}. Facing the demand to implement compliance with regulations such as EAA, enterprises conduct regulatory requirements engineering (\textit{regulatory RE}), deriving software requirements from intentionally abstract regulatory texts. Our study focuses on \emph{Regulatory Impact Analysis} (RIA). Based on our results, we define RIA as a regulatory RE activity conducted at an enterprise-wide level to coordinate the implementation of regulation throughout the enterprise and develop internal requirements (policies) applicable to multiple products. RIA also exists in policy-making, aiming to understand whether regulations had the intended impact or require improvement~\cite{ellig2018and}. In the context of regulatory RE, we take the perspective of the persons affected by the regulation, not the regulators.

\section{Related work}\label{sec:relatedWork}
In this section we introduce three research streams that cover some of the aspects of regulatory RE in large enterprises in a fragmented way.

\subsection{Large-scale (regulatory) requirements engineering}
Existing studies identify large-scale requirements engineering as requirements engineering dealing with around 1000 requirements within a project or product line~\cite{regnell2008can}. To the authors' best knowledge, no studies consider the size of organizations, number of products, or the number of stakeholders as factors defining the large scale of RE. The extensive body of literature on regulatory RE~\cite{mubarkoot2023software} did not pay significant attention to the empirical research of regulatory RE in large enterprises. The case study by Usman et al.~\cite{usman2020compliance} is the only one known to the authors specifically focusing on this topic. Usman et al.~\cite{usman2020compliance} found that regulatory RE in large enterprises is conducted at five levels. This study identified that the central unit is responsible for maintaining compliance requirements and developing the design rules and guidelines for the SIPS teams. However, the study did not cover the role of the central unit or its coordination with other levels in detail. Other empirical studies~\cite{klymenko2022understanding, amaral2023nlp, kosenkov2024developing} pointed out the importance and challenges of engineering-legal interaction and implicit nature of the legal knowledge in required the context of regulatory compliance. However, these studies did not focus on the structure of such interaction or its broader enterprise context.

\subsection{Team coordination and knowledge management research}
Another relevant track of research relevant is the research on SIPS team coordination and knowledge management. Kasauli et al.~\cite{kasauli2020charting} identified that teams applying different specialized methods (e.g., agile, waterfall) and belonging to different disciplines (e.g., hardware, software) often need to collaborate on the same product~\cite{kasauli2020charting}. Boundary objects (artifacts in which knowledge is manifested~\cite{wohlrab2019boundary}) can then be used for interaction, knowledge sharing between teams and the integration of the work of such teams~\cite{wohlrab2019boundary}. Kasauli et al.~\cite{kasauli2020charting} found that standards, regulations, and safety assurance cases are such typical boundary objects used to ensure common understanding and compliance. Studies in this research track mention regulatory compliance among the concerns in horizontal inter-team coordination. However, these studies did not focus on regulatory RE specifically.

\subsection{Regulatory compliance in (scaled) agile methodologies}
Large enterprises use agile methodologies (e.g., Scaled Agile Framework (SAFe), LeSS) to scale agile practices throughout the enterprise. Existing studies mention that organization-wide compliance processes or regulatory changes are among the challenges to applying scaled agile methods~\cite{conboy2019implementing}. SAFe includes continuous mitigation of compliance concerns as one of its ten practices of enterprise solution delivery~\cite{safe2024compliance}. Some studies suggested approaches to incorporating regulatory compliance into scaled agile (e.g.,~\cite{moyon2020integrate, poth2020systematic}). Multiple studies have explored the application of agile methodologies in highly regulated industries (e.g., medical devices~\cite{stirbu2018towards}). However, these studies typically focus on a compliance on the level of a single SIPS team rather than on an enterprise level. Other studies (e.g.~\cite{nagele2023current}) mentioned the conflict between governance procedures with the autonomy of agile teams, however have not explored the governance demands. 
In large enterprises, executive management is one important stakeholder group that requires a backbone and maintains a certain level of governance in the enterprise~\cite{kasauli2020charting}. However, neither existing practical frameworks nor studies in this track focused on regulatory RE at the enterprise level considering executive management <<demands>>.

\section{Methodology}\label{sec:methodology}
To answer our research questions, the first three authors conducted three group semi-structured exploratory interviews with nine experts involved in the RIA process in three large enterprises (E1, E2, E3). We have followed existing guidelines by Linaaker et al.~\cite{linaaker2015guidelines} and Runeson et al.~\cite{runeson2009guidelines} to conduct the interviews.

\textbf{Participant Selection} To select the interviewees, we have sent out a call for participation to large enterprises developing SIPS for their customers and, hence, potentially falling under the EAA. Our call was directed towards external and internal roles with technical, legal, and accessibility competence involved in the RIA execution. We used purposive and convenience sampling and selected participants who corresponded to the selection criteria, responded to our call, and were available. In each of the three enterprises that responded (see Table~\ref{tab:enterprises} for details), three persons were selected for the interviews, with the overall number of RIA initiative participants in each enterprise varying between six and eleven.

\textbf{Data Collection} We organized interviews in sessions to ensure data validity. In E1, we conducted one interview session and did not conduct any follow-up interviews as structured and sufficient data was collected. We produced a summary of our findings and requested the interviewees to provide written feedback. However, no feedback was received. In E2 and E3, we conducted initial interview sessions to collect the data. After that, we analyzed the interviews to produce a summary of the collected data built around the main themes of the interviews and formulate intermediary conclusions. Next, we conducted follow-up interview sessions to validate the data and discuss our intermediary conclusions. After conducting the interviews in E3, we observed that we had reached saturation as no new significant data about the RIA process across different enterprises emerged. All sessions were conducted remotely via Microsoft Teams and were recorded. The duration of the interview sessions varied between 1 and 2.3 hours (see Table~\ref{tab:enterprises}).
We started the initial interview sessions with general questions about the RIA process and transitioned to concrete questions focusing on RIA activities and the characteristics of these activities. Our questions and discussions were built around the following themes: general characteristics of RIA, stakeholders, and roles participating in RIA, the interaction between the roles and stakeholders, RIA activities, goals of RIA overall and concrete activities, artifacts used as inputs and outputs in the activities, required knowledge, and challenges encountered (see  ~\href{https://doi.org/10.5281/zenodo.13747866}{the list of the questions here (DOI: 10.5281/zenodo.13747866)}.

\textbf{Data Analysis and Synthesis}
To analyze the data collected during the initial interview sessions, the first author transcribed the interviews and conducted a thematic analysis focused on the main themes of the interviews. During the thematic analysis, we did not identify any additional themes beyond the ones predefined for questions and hence used code "other" to collect additional observations. The second author conducted a review of the transcripts (without coding) in parallel, watched the recordings, and made notes. To analyze the information collected during the follow-up group interviews, the paper's first and second authors listened to the interviews' recordings and made notes. Due to the sensitive nature of the information about the ways of working collected during the interviews, we can not disclose the raw interview data.

\vspace{-.9cm}
\begin{table}
\begin{center}
\setlength{\tabcolsep}{12pt}
\caption{Overview of enterprises involved in the study}
\label{tab:enterprises}
\begin{adjustbox}{width=0.9\textwidth}
\begin{tabular}{l c c c} 
 \toprule
  & \textbf{Enterprise 1 (E1)} & \textbf{Enterprise 2 (E2)} & \textbf{Enterprise 3 (E3)} \\
 \midrule
 \textbf{Industry} & Telecommunications & Finances & Finances\\
 \midrule
 \textbf{Employees} & >15,000 & >10,000 & >15,000\\
 \midrule
 \textbf{Countries present in} & 7 & 7 & 4(8)\\
 \midrule
 \textbf{Agile methodology} & SAFe & SAFe & SAFe\\
 \midrule
 & Enterprise architect&Accessibility expert&Customer experience expert\\
 
 \textbf{Interviewees} & Enterprise architect&Technical expert&UI design expert\\
 
  & Legal expert& Legal expert&Accesibility expert\\
 \midrule
\textbf{Interviews duration} & 2.3 hours & 1.5 + 1.5 hours & 1 + 1.3 hours \\
 \bottomrule
\end{tabular}
\end{adjustbox}
\end{center}
\end{table}
\vspace{-1.3cm}

\section{Results}\label{sec:results}
This section reports the results of the interviews by grouping them according to the main themes of the interviews. We then take the perspectives of the research questions in the discussion of the results (Sec.~\ref{sec:discussion}).

\subsection{RIA general characteristics}
In all three enterprises, RIA was executed as the first step in processing the EAA after it was identified as applicable to an enterprise. Notably, in all three enterprises, RIA was initiated by stakeholders not belonging to legal or compliance functions. During the interviews, enterprises were at some stage of the RIA process or recently completed it and were conducting follow-up activities (e.g., cost estimates). RIA was executed by a cross-functional group of experts (RIA group, see Section~\ref{sec:stakeholders} for details) created by executive management or a central function. The number of participants in the RIA group varied between six and eleven participants across three enterprises. The enterprises had formally established a process for RIA in all three cases. The developers/owners of the process were enterprise architecture, governance, or compliance functions. In E1, the process included seven activities; in E2, it consisted of four activities; and in E3, it included five activities (see generalized information about activities in Section~\ref{sec:activities}). In E2 and E3, the interviewees reported that the process was under redesign because its existing version was too high-level and could not be effectively applied. RIA for the EAA was characterized as repeatable (potentially requiring execution for EAA in the future). In all three cases, RIA was also executed for other regulations (e.g., GDPR, DORA).

\subsection{RIA group roles \& RIA stakeholders}\label{sec:stakeholders}
We have identified roles involved in RIA execution and stakeholders whose interests were impacted by RIA.
The RIA groups included technical experts, representatives of enterprise architecture or governance, legal or compliance functions, internal and/or external accessibility experts, and representatives of geographical or product OUs. In E1 and E2, enterprise management and EAA implementation sponsor (appointed as a representative by executive management) were identified as the main stakeholders of the RIA process. In E3, the compliance function was the stakeholder to which the RIA group reported.

\begin{colorb}
\small
\textit{The impact analysis is not aimed at product managers; it's aimed at the higher management or sponsor to help implement [EAA]. The main purpose is not to say to product management this is what you need to do.} (E1)
\end{colorb}

The RIA groups interacted with multiple support functions, such as the central legal/compliance function, the user interface component organization, risk management, and external experts. RIA groups also delegated or coordinated some of the RIA tasks with responsible roles in different national OUs, product OUs or SIPS teams, and suppliers.

\subsection{Goals}\label{sec:goals}
We have identified four core goals of RIA in all three enterprises. We elaborate on these goals next.

\textbf{Aggregated view on EAA impact}. The core goal of the RIA process is the provision of the enterprise-wide aggregated view (across different national and product OUs) on EAA impact and aggregated view on compliance implementation. Such an aggregated view primarily serves the purpose of supporting executive management in their decision-making process. Interviewees from E1 emphasized that identifying the aggregated impact of all the applicable regulations is a goal of RIA but also a challenge that needs to be addressed.
Some of the additional reasons for an aggregated view were compliance risk management (E1) and the selection of the ``harmonized'' (similar across the enterprise) level of compliance (E1, E2).
In E1, the interviewees reported that the decentralized RIA process, in which full responsibility is delegated to business and product owners, could not assure such an aggregated view.

\textbf{Enabling efficient EAA compliance}. Another goal of RIA was to facilitate effectiveness and efficiency in implementing compliance. This included developing a structured EAA implementation strategy and approach to resource allocation (E2, E3). Also, the achievement of this goal included identifying the responsible roles (E3).
Scaling compliance measures (audits, trainings in E1) were important for enterprises to optimize expenditures.
In E2, the interviewees pointed out that full delegation of RIA and compliance implementation to SIPS teams can result in only partial compliance and additional expenses.
\begin{colorb}
\small
\textit{We could leave it to [SIPS teams], ask them to develop according to the EAA. Then check and find out that we are 75\% compliant and 25\% needs to be redone. But that would be a too expensive way to do it.} (E2)
\end{colorb}

\textbf{Facilitating planning}. Another goal of RIA in all three cases was enabling the planning of EAA compliance implementation (e.g., time and costs).
In E1, the interviewees emphasized that RIA was also directed towards ensuring that product managers address the EAA in their planning. According to the interviewees, such planning delegation was required because it was impossible to conduct effective cost estimates on the same high level at which RIA was conducted.

\textbf{Assuring EAA compliance governance}. All three enterprises emphasized that RIA also needed to help with the governance and/or maintenance of EAA compliance in the future. Some activities directed towards this goal were creating governance and support structures (E2, E3) and developing policies. In E2, the interviewees emphasized that it is also essential to allocate responsibility to assure ownership and compliance maintenance (E2).

\subsection{Activities}\label{sec:activities}
\begin{figure}
    \centering
    \includegraphics[width=0.75\textwidth]{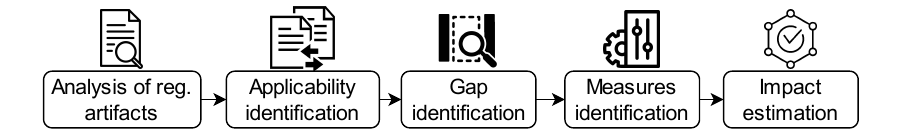}
    \caption{Main tasks of RIA as synthesized from three enterprises.}
    \label{fig:riaTasks}
\end{figure}

The number, naming, and execution of RIA activities varied in all three enterprises. However, we identified five core types of activities executed in all three companies. Figure~\ref{fig:riaTasks} provides an aggregated overview of these main types of RIA activities, originating from synthesizing the three cases. We describe each of them next.
\textit{Analysis of regulatory artifacts} was executed by the legal experts from the RIA group to produce basic analysis of the EAA (e.g., summarize it, categorize EAA norms).
\textit{EAA applicability analysis} was conducted to identify where in the enterprises EAA applies. First, legal experts iteratively collaborated with other participants of the RIA group possessing organizational expertise to identify impacted OUs. Secondly, the RIA group delegated a detailed EAA applicability analysis to the responsible roles in OUs (E2, E3) or conducted interviews with them (E1).
The \textit{Gap identification} activity was directed towards identifying the difference between the normative required state of accessibility and its current state. It was also conducted on both the enterprise level by RIA and on the level of OUs by responsible roles.
During the compliance \textit{Measures identification}, the RIA group was mainly focused on enterprise-wide measures (e.g., policies, training). In this activity, the RIA group analyzed measures planned on the level of OUs and SIPS teams.
\textit{Impact assessment} was the activity concluding RIA and identifying the degree or approximate cost of the measures for EAA implementation both on the enterprise and OUs level. Practically all activities in the RIA process required cross-functional interaction within the RIA team and vertical interaction between the RIA team and OUs in some form. For example, in E1 the results of the analysis of regulatory artifacts were collaboratively adapted to make them useful for subsequent RIA activities and executive management.

\vspace{-.5cm}
\vspace{-1.1cm}

\subsection{Artifacts}\label{sec:artifacts}
We identified three main types of artifacts used in RIA: (1) external regulatory artifacts processed to conduct RIA, (2) internal interpretation artifacts used and/or produced while processing regulatory artifacts, (3) compliance artifacts specifying RIA results and intended for supporting the implementation of RIA results and/or EAA compliance in the future.

\textbf{Regulatory artifacts.} Despite this study being focused on RIA for the EAA, we have identified several other regulations that enterprises process along with the EAA: national accessibility laws, legislative preparatory work for national accessibility legislation, standards and guidelines on accessibility (e.g., WCAG 2.0, standard EN 301549). In E1, the legal expert reported that there are multiple levels of regulatory acts (e.g., European, national) and preparatory work ("doctrine") that are processed to interpret EAA.

\textbf{Interpretation artifacts.} In all three enterprises, we identified two main types of artifacts used in interpreting regulatory artifacts, i.e. their application to a particular case. These are (1) input ``case artifacts'' with information about enterprise,  OUs, and SIPS, (2) communication artifacts used to exchange information, and (3) output artifacts with the results of the interpretation of regulatory artifacts. Templates (usually in Word or Excel format) were used to exchange information between the RIA group and other stakeholders. In E3, the interviewees mentioned using a couple of templates in the RIA process for sending information to legal functions. Interviewees considered these templates too high-level for RIA purposes and only served legal function needs. Hence, they were looking for an opportunity to restructure this template.

\begin{colorb}
\small
\textit{It is challenging to find a good format and not fill in a template because you must send it to compliance.} (E3)
\end{colorb}

The artifacts representing the results of RIA were as follows: legal analysis of EAA (focused on different articles) (E1, E2) or summary of its scope (E3), interpretation of EAA across countries and/or mapping of differences (E1, E2, E3), mapping of EAA norms and impacted OUs and/or products (E1, E2, E3), results of gap assessment in OUs and on the enterprise level (E1, E2, E3).

\begin{colorb}
\small
\textit{We bounced the table off a few times because of legal way of describing things. We made it to illustrate the requirements in a way practical for non-lawyers.}(E1)
\end{colorb}

\textbf{Compliance artifacts.} In all three enterprises, new artifacts were introduced, or existing artifacts were changed to specify RIA results and/or support the implementation of EAA.
Some of the new compliance artifacts were guidelines, standards, policies, instructions (E1, E2, E3), repository/framework of compliant components (E2, E3), implementation plans for OUs (E2, E3), plan for educational activities (E2), risk assessment (E2), and public documentation on EAA compliance (E2).
Compliance artifacts were closely related. For example, in E2 and E3, the interviewees mentioned that the availability of EAA-compliant components or patterns per se does not guarantee compliance implementation; rather, it also depends on the training.
\begin{colorb}
\small
\textit{If the team uses [a framework of components], the central organization guarantees that these components are compliant. That does not help because you need to understand how to get the components together.} (E2)
\end{colorb}

The existing artifacts in which changes were required were as follows: ways of working (E1), existing design guidelines to introduce ``accessibility by design'' (E1), testing processes (E1), roles and responsibilities (E2), requirements and development processes (E2), and a definition of done (E2).

\subsection{Knowledge and expertise}\label{sec:knowledge}
The main types of knowledge and expertise involved in RIA were legal knowledge (applicable regulations, systematic understanding of legal analysis and interpretation), organizational knowledge (structure of enterprises, viability of concrete products, risks related knowledge), technical knowledge (knowledge of concrete products and portfolio, measures required for EAA compliance and estimation of their costs).
The RIA process relies both on documented and undocumented knowledge. For example, in E1, RIA was heavily based on documented enterprise architecture artifacts, while in the other cases, RIA relied on the knowledge of the group participants or the delegation of activities to OUs.
We discovered that the legal knowledge used in the RIA process was implicit. Despite our efforts to elucidate the specific legal knowledge and its application, the legal experts in E1 and E2 were unable to clearly explain their methods of legal analysis.
\begin{colorb}
\small
    \textit{- How did you know that this legal norm applies?\\
    - Because I have been working in [industry] regulation for many years and I know it outside out.} (E1)
\end{colorb}
In E3, the interviewees emphasized that interaction with experts with corresponding knowledge plays a key role in RIA.

\begin{colorb}
\small
\textit{It is not the case of you know it all when it comes to big organizations. It's rather that you have contacts and can reach out to people in the different units.}(E3)
\end{colorb}

\subsection{Challenges}\label{sec:challenges}
We have identified eight challenges (Ch1-8) to RIA that interviewees mentioned explicitly or which we identified and discussed with interviewees in follow-up interview sessions.

\textbf{Ch1: Accommodating Agile and waterfall ways of working.} In all three enterprises, the interviewees reported that they experienced challenges in executing waterfall/plan-based RIA in Agile organizations. In E3 this resulted in the application of hybrid approaches in RIA execution.

\begin{colorb}
\small
\textit{We are working bottom-up, but also top-down to make sure that we get buy-in, that we motivate and give the "Why" [to SIPS teams]. We try to find a balance on how to meet the demands on different levels.} (E3)
\end{colorb}


In E2 and E3, the interviewees emphasized that EAA compliance should be implemented efficiently and in a way suitable for SIPS teams.
\begin{colorb}
\small
\textit{We are decentralized and need to assure that everybody grows and takes responsibility. [We] need to make everybody feel comfortable working as they want.}(E2)
\end{colorb}
In E2, the interviewees mentioned that some regulations (e.g., MiFID) are in the area of responsibility of the ``business side'' because they are related to particular products only. For such regulations, RIA is not required.

\textbf{Ch2: Implicit and undocumented knowledge.} The interviewees reported a few challenges related to implicit and/or undocumented knowledge (see Section~\ref{sec:knowledge}). The main challenge was that only some information on the previous implementation of accessibility regulations was documented and reusable. One of the implications of undocumented organizational knowledge was the risk of missing the OUs or SIPS teams responsible for SIPS components (E2). In E1, the legal expert also reported that delegating legal analysis within the RIA to legal experts without experience in the corresponding industry is difficult.

\textbf{Ch3: Multiplicity of regulations.} The interviewees mentioned that multiple accessibility-related regulatory artifacts need to be processed (see also Section~\ref{sec:artifacts}).
In E3, the interviewees mentioned the implementation of concurrently applicable regulations as challenging:
\begin{colorb}
\small
\textit{There are a lot of regulations and things we cater for like security. That's a challenge that we need to make sure that everything works in a good way.}(E3)
\end{colorb}

The multiplicity of regulations also results in conflicts between regulations (e.g., accessibility and security requirements (E3)).

\textbf{Ch4: Interpretation and implementation consistency.} In the follow-up interview sessions in E2 and E3, we discussed the challenge of consistency of EAA interpretation. We found out that, while the RIA group was conducting an initial interpretation of EAA, further delegation of RIA tasks or future EAA implementation could lead to new interpretations by other roles. In E3, one of the interviewees stated that the interpretation of regulations is also a part of SIPS development. The interviewees confirmed that this imposes a challenge in interpreting consistently, and there are currently no approaches to address that.

\textbf{Ch5: Practical applicability of RIA.} In E1 and E2, the interviewees explicitly mentioned the challenge of making RIA results practically applicable for SIPS teams. In E1, this challenge was mentioned in the context of the development of EAA interpretation artifacts, and in E2, it was related to the communication of RIA results to SIPS teams.

\begin{colorb}
\small
\textit{That's more of a problem to get the developing organization to understand how to implement compliance.}(E2)
\end{colorb}

\textbf{Ch6: Demand for expert knowledge.} In all three enterprises RIA involved four categories of expertise: technical expertise about SIPS, legal expertise, accessibility expertise, and organizational expertise. In E3, the interviewees mentioned the demand for different types of expertise as a significant challenge.

\begin{colorb}
\small
\textit{There are developers, designers, testers, who need to understand and know everything. That is a huge challenge because how should they be able to be experts on accessibility, on GDPR, on sustainability, etc.} (E3)
\end{colorb}

Expert involvement was also required to apply documented knowledge (e.g., organizational knowledge to identify impacted OUs in E1).

\textbf{Ch7: Measuring compliance.} In E2 and E3, identifying how to measure compliance was explicitly described as challenging. Measurement of EAA compliance was seen as essential to: (1) verify compliance, (2) sustain it in the future, (3) making decisions about the feasible level of compliance.
In E3, the interviewees elaborated that no concrete measurement or tool is applicable for accessibility compliance measurement.

\textbf{Ch8: Sustaining compliance.} The challenge of sustaining the RIA results and future EAA compliance was related to the need to establish new roles and/or structures, scaling of organizations, and personnel turnover (E3).

\begin{colorb}
\small
\textit{We cannot run to the different areas or teams and reorganize. We're trying to find roles that they need to establish so we can reach out to one person. That role is responsible for providing information to their organization and giving back information to us.} (E3)
\end{colorb}

In E2 and E3, the interviewees emphasized that due to the absence of approaches to sustain compliance, their enterprises have not secured the  accessibility compliance  implemented before.

\begin{colorb}
\small
\textit{We were good at [accessibility] 15-20 years ago. Things slip away and disappear if you do not set the structure for continuous work. It's one thing to implement, but it's a completely different thing to run it, and make it non-personal, and future proof it.} (E3)
\end{colorb}

\vspace{-.7cm}
\begin{figure}
    \centering
    \includegraphics[width=0.8\textwidth]{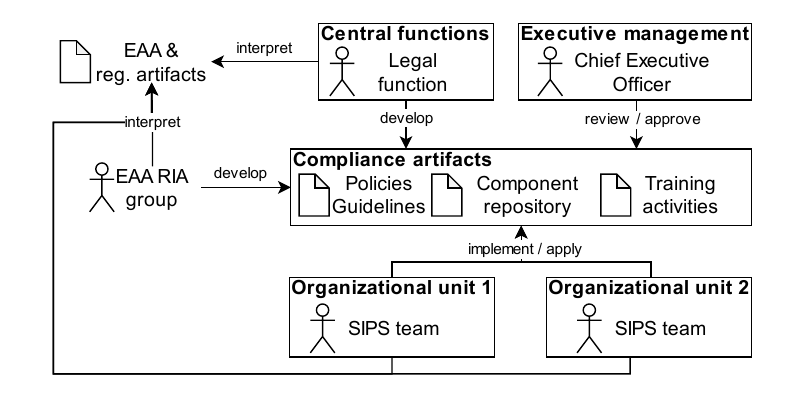}
    \caption{Visualization of coordination in the process of RIA. For complete information roles, artifacts see corresponding sections.}
    \label{fig:coordination}
\end{figure}
\vspace{-.8cm}

\section{Discussion}\label{sec:discussion}

\begin{colora}
    \textbf{Answer to RQ1:} 
    Executive management is a special group of stakeholders in the process of regulatory RE. In order to address the executive management needs in large enterprises, RIA is executed through the coordination of multiple roles and stakeholders across different functions (cross-functional coordination) and enterprise levels (vertical coordination) (see Figure~\ref{fig:coordination} for coordination overview).
\end{colora}

Specially created cross-functional RIA groups are key in executing such coordination. The most important \textit{cross-functional coordination} within the RIA group happens between the technical, legal, accessibility, and organizational experts. The most important vertical coordination happens between central functions on the one side and OUs and their SIPS teams on the other side. Based on our observations, we denote \textit{coordination} in regulatory RE as an interaction between roles belonging to different functions and/or enterprise levels in which they depend on the results of each other for the achievement of a common goal. Coordination is inherent to all of the RIA activities and plays a key role to enable decisions in RIA and EAA compliance (e.g., minimal and "harmonized" level of EAA compliance). To the best of our knowledge, no studies have specifically explored and defined coordination in software engineering from the cross-functional perspective. We suggest that cross-functional and vertical coordination introduce different challenges in comparison to vertical inter-team coordination, which received more attention in existing studies.

We observed that regulatory RE establishes the basis for the EAA compliance governance (e.g., level of compliance, compliance measurement) common for both SIPS teams and legal functions. Hence, we suggest that regulatory RE in settings other than large enterprises (e.g., small and medium enterprises), should also address business, organizational, risk management concerns to ensure effective EAA compliance governance. Future research on cross-functional and vertical coordination in regulatory RE is required to identify the generalizability of our results to small and medium enterprises.

\begin{colora}
    \textbf{Answer to RQ2:} We identified eight main challenges to RIA: (1) accommodation of agile and waterfall ways of working, (2) implicit and undocumented knowledge, (3) multiplicity of regulations, (4) interpretation and implementation consistency, (5) practical applicability of RIA, (6) demand for expert knowledge, (7) measurement of compliance, (8) maintenance of compliance. The challenges are directly or indirectly related to coordination and knowledge management.
\end{colora}

The identified challenges are closely connected. As SIPS teams cannot handle all types of requirements deriving from multiple regulations independently (\textit{Ch3}), there is an evident demand to involve corresponding experts (\textit{Ch6}). Such emerging coordination comes with challenges of accommodation of agile ways of working in OUs and SIPS teams and waterfall-type ways of working of legal and central functions (\textit{Ch1}) and making the results of legal analysis actionable for OUs and SIPS teams (\textit{Ch5}). RIA and regulatory RE inputs and outputs are knowledge-intensive and hence encounter the challenge of implicit and undocumented knowledge (\textit{Ch2}). For example, our results also confirm that regulatory RE involves the processing of auxiliary regulatory artifacts identified according to the legal knowledge that is implicit to non-legal roles~\cite{amaral2023nlp}. Also, the absence of well-documented RIA artifacts can lead to interpretation and implementation inconsistencies in different OUs and SIPS teams (\textit{Ch4}). Finally, vertical coordination is essential to ensure appropriate measurement of compliance (\textit{Ch7}) when responsibility for compliance implementation is distributed across multiple SIPS teams. Our results indicate that in practical settings, regulatory RE legal expertise is indispensable for multiple purposes (e.g., interpreting the numerous regulatory artifacts). Current regulatory RE studies often overlook the relationships between regulations and fail to discuss the legal knowledge needed to understand these connections. We propose that future regulatory RE research should consider the necessity of legal expertise and knowledge management. Our study also discovers that in practice regulations such as EAA are addressed both on the enterprise-wide level (in enterprise policies or other compliance artifacts) and on the level of SIPS teams. Herewith, we suggest that the challenges Ch2, Ch4, Ch7, and Ch8 require knowledge management practices. Knowledge management is essential, for example, to avoid conflicts between interpretation on the enterprise level and in SIPS teams (\textit{Ch4}) and to sustain compliance (\textit{Ch8}). Knowledge management in regulatory RE also can be essential for small and medium enterprises in which it should support the consistent interpretation of regulations and understanding of business, organizational or other relevant aspects. throughout time.

\begin{colora}
    \textbf{Answer to RQ3:} Large enterprises address the cross-functional interaction in the RIA process by (1) introducing cross-functional RIA groups and (2) producing and tailoring artifacts.
\end{colora}
We discovered that RIA as a part of regulatory RE includes coordination not only between engineering and legal roles, but also roles possessing organizational and accessibility domain knowledge. The two types of the artifacts used for cross-functional coordination are: (1) interpretation artifacts used in the RIA group (e.g., legal analysis results) and (2) compliance artifacts (e.g., policies). However, creating and tailoring such artifacts is not systematic and requires improvement.

\section{Threats to validity}
\textbf{Internal validity} To assure the validity of the collected data and validate our intermediary conclusions, we conducted follow-up interview sessions in E2 and E3, while in E1, provided data was structured enough to skip such follow-up. To ensure the consistency of the interview process, the first three authors conducted the interviews, paying attention to covering all pre-defined topics and questions (SEE). The interviewees received a detailed description of the study scope beforehand, and the interview questions were internally reviewed. We used group interviews to mitigate bias on the side of interviewees and facilitate discussion in case of differences in opinions.
We predefined our focus themes to avoid bias in the qualitative coding process and conducted follow-up interview sessions. The first two authors of the paper also reviewed transcripts. For our analysis and conclusions, we mainly focused on the results relevant to all three enterprises and the results that answered our research questions. 

\textbf{External validity} As we have focused on the RIA as a regulatory RE activity in large enterprises, we involved enterprises from two different industries. We did not focus on selecting companies using different scaled agile methodologies as this was out of the scope of our initial interview but emerged as a side result during the interviews.
We argue that we have reached sufficient saturation as no new significant results emerged during the interviews with interviewees from the same enterprise after the follow-up interviews, and no new significant differences in the RIA process across enterprises emerged after the analysis of the interviews conducted with the third enterprise. Additionally, all three enterprises had basic documentation for the RIA process, establishing a common understanding among the interviewees. Consequently, we deemed the involvement of three interviewees from each enterprise sufficient.
We believe our results are to a certain degree generalizable to other regulations, as the RIA process was pre-established and mandatory for any regulation in all three enterprises. Interviews explicitly indicated that RIA (despite being abstract) was applied to various regulations (e.g., GDPR, DORA). We suggest that any potential differences in the execution of RIA across regulations can be primarily related to the content of the regulations (e.g., legal concepts and their relationships). At the same time, the core RIA tasks identified in this study remain the same for any regulation~\ref{fig:riaTasks}.


\section{Conclusion and future work}\label{sec:conclusion}
Large enterprises face a need to implement compliance with the growing amount of regulations. Our exploratory results show that an increasing number of large enterprises are trying to account for regulations by conducting regulatory impact analysis as their first regulatory RE activity to implement regulations. The results of our study suggest that RIA and regulatory RE processes in large enterprises involve multiple cross-functional concerns and stakeholders. In particular, they involve enterprise stakeholders whose interests on the enterprise-wide level need to be considered by development teams. Regulatory RE in large enterprises requires cross-functional coordination in which legal, enterprise, and technical knowledge is applied by corresponding roles. Our study's essential finding is that interpretation and compliance artifacts produced in the RIA process serve as boundary objects to communicate enterprise stakeholders' requirements to organizational units and teams developing software-intensive products and services. We suggest that appropriate management of such artifacts can address the main challenges to cross-functional coordination and knowledge management.

We plan to extend the results of our interview study and conduct in-depth case studies that would better capture the perspective of development teams themselves in regulatory RE. We also plan to explore the regulatory and compliance artifacts in detail and incorporate them into regulatory RE currently under development and evaluation~\cite{kosenkov2024developing}.

 \bibliographystyle{splncs04}
 \bibliography{bibliography}

\end{document}